\documentclass[aps,prd,nofootinbib,10pt]{revtex4-1}
\usepackage[colorlinks=true,linkcolor=black,citecolor=blue,urlcolor=black]{hyperref}
\usepackage{natbib}
\usepackage{amssymb,amsmath}
\usepackage[utf8x]{inputenc}
\usepackage{graphicx}
\usepackage{subfig}
\usepackage{caption}
\usepackage[export]{adjustbox}
\usepackage{cancel,float}
\usepackage{color}
\DeclareGraphicsExtensions{.png,.eps,.eps}
\usepackage{axodraw}
\usepackage{graphicx}
\usepackage{color}

\def\ev{{\rm eV}}

\def\tev{{\rm TeV}}

\renewcommand{\baselinestretch}{1.5}

\def\beq{\begin{equation}}
\def\eeq{\end{equation}}
\def\barr{\begin{eqnarray}}
\def\earr{\end{eqnarray}}
\def\dis{\displaystyle}

\def\beq{\begin{equation}}
\def\eeq{\end{equation}}
\def\bea{\begin{eqnarray}}
\def\eea{\end{eqnarray}}
\def\nn{\nonumber}

\def\sss{\scriptscriptstyle}

\def\roughly#1{\mathrel{\raise.3ex\hbox
{$#1$\kern-.75em\lower1ex\hbox{$\sim$}}}}
\def\lsim{\roughly<}

\def\sla#1{\raise.15ex\hbox{$/$}\kern-.57em #1}
\def\bra#1{\left\langle #1\right|}
\def\ket#1{\left| #1\right\rangle}

\def\bd{B_d^0}

\def\ANPq{{\cal A}^q}

\def\ApNPqph{{\cal A}^{\prime,q} e^{i \Phi'_q}}
\def\ApNPCqph{{\cal A}^{\prime {C}, q} e^{i \Phi_q^{\prime C}}}
\def\ApNPCuph{{\cal A}^{\prime {C}, u} e^{i \Phi_u^{\prime C}}}
\def\ApNPCdph{{\cal A}^{\prime {C}, d} e^{i \Phi_d^{\prime C}}}
\def\pewcp{P_{EW}^{\prime C}}
\def\pewp{P'_{EW}}
\def\pewcpnp{P_{EW, NP}^{\prime C}}
\def\pewpnp{P'_{EW, NP}}
\def\ApNPuph{{\cal A}^{\prime,u} e^{i \Phi'_u}}
\def\ApNPdph{{\cal A}^{\prime,d} e^{i \Phi'_d}}
\def\ApNPcomb{{\cal A}^{\prime, comb} e^{i \Phi'}}
\def\btopik{B \to \pi K}
\def\bsmumu{b \to s \mu^+ \mu^-}
\def\bsll{b \to s \ell^+ \ell^-}

\def\s{\sqrt{2}}

\def\bd{B^0}

\def\bsee{b \to s e^+ e^-}
\def\bsmumu{b \to s \mu^+ \mu^-}

\def\bsee{b \to s e^+ e^-}
\def\bsll{b \to s \ell^+ \ell^-}

\def\bctaunu{b \to c \tau^- {\bar\nu}}

\def \cB{{\cal B}}
\def \SM{{\rm SM}}
\def \NP{{\rm NP}}

\def\RK{{R_K}}

\def\RKstar{{R_K^*}}

\begin{document}
\title{\bf \boldmath Unified explanation of $\bsmumu$ anomalies, neutrino masses and $B\rightarrow \pi K$ puzzle}

\author{Alakabha Datta}
\email{datta@phy.olemiss.edu}
\affiliation{Department of Physics and Astronomy, 108 Lewis Hall, University of Mississippi, Oxford, Mississippi 38677-1848, USA}
\author{Divya Sachdeva}
\email{divyasachdeva951@gmail.com}
\affiliation{Department of Physics and Astrophysics, University of Delhi, Delhi 110 007, India}
\author{John Waite}
\email{jvwaite@go.olemiss.edu}
\affiliation{Department of Physics and Astronomy, 108 Lewis Hall, University of Mississippi, Oxford, MS 38677-1848, USA}

\begin{flushright}
 UMISS-HEP-2019-01 \\
\end{flushright}

\begin{abstract} Anomalies in semileptonic $B$ decays  could indicate new physics beyond the standard model(SM).
There is an older puzzle in nonleptonic $B \to \pi K$ decays. The new particles, leptoquarks and diquarks, 
required to solve the semileptonic and the nonleptonic puzzles can also generate neutrino masses and mixing
at loop level. We show that a consistent framework to explain the $B$ anomalies and the neutrino masses is possible and we make predictions
for certain rare nonleptonic $B$ decays.
\end{abstract}  

\maketitle

\section{Introduction}
Searching for beyond the SM (BSM) physics has been the primary focus of the high energy community.
Rare $B$ decays have been widely studied to look for BSM effects. Because these decays get small SM contributions, new physics (NP)
can compete with the SM and produce deviations from SM predictions. Over the last few years measurements in certain
$B$ decays have shown deviations from the SM. These deviations are observed in two groups$-$ in charged current (CC)
processes mediated by the $\bctaunu$ tansitions and in the neutral current (NC) processes mediated by $\bsll$ transition
with $\ell= \mu, e$. We will focus here on the NC anomalies although
it is possible that the CC and the NC anomalies are related \cite{Bhattacharya:2014wla} but we will not explore
that possibility here.


Let us start with the $\bsll$ decays which are fertile grounds to look for new physics effects \cite{Alok:2010zd, Alok:2011gv}. 
In $\bsmumu$ transitions
 there are discrepancies with the SM in a number of observables in $B
\to K^* \mu^+\mu^-$ \cite{Aaij:2013qta, Aaij:2015oid, Abdesselam:2016llu,
ATLAS:2017dlm, CMS:2017ivg} and $B_s^0 \to \phi \mu^+ \mu^-$
\cite{Aaij:2013aln, Aaij:2015esa}. 

There are also measurements that are different from the SM expectations that involve ratios of $\bsmumu$ and $\bsee$ transitions.
These measured quantities are tests of lepton universality violation (LUV) and are defined as
 $R_K \equiv \cB(B^+ \to K^+
\mu^+ \mu^-)/{\cal B}(B^+ \to K^+ e^+ e^-)$ \cite{Aaij:2014ora, Aaij:2019wad} and $R_{K^*}
\equiv {\cal B}(B^0 \to K^{*0} \mu^+ \mu^-)/\cB(B^0 \to K^{*0} e^+
e^-)$ \cite{Aaij:2017vbb, Abdesselam:2019wac}.

While the discrepancies in $\bsmumu$ can be understood with  lepton universal  new physics\cite{Datta:2013kja}, hints of LUV  in $\RK$ and $\RKstar$ require
NP that couple differently to the lepton generations.  A well-studied scenario is to assume NP coupling dominantly
to the muons though NP coupling to electrons
is not ruled out \cite{Datta:2017ezo, Datta:2019zca}.  
The $\bsmumu$ transitions are defined via an effective
Hamiltonian with vector and axial vector operators:
\bea
H_{\rm eff} &=& - \frac{\alpha G_F}{\s \pi} V_{tb} V_{ts*}
      \sum_{a = 9,10} ( C_a O_a + C'_a O'_a ) ~, \nn\\
O_{9(10)} &=& [ {\bar s} \gamma_\mu P_L b ] [ {\bar\mu} \gamma^\mu (\gamma_5) \mu ] ~,
\label{Heff}
\eea
where the $V_{ij}$ are elements of the Cabibbo-Kobayashi-Maskawa (CKM)
matrix and the primed operators are obtained by replacing $L$ with
$R$. It is assumed Wilson coefficients (WCs) include both the SM and NP
contributions: $C_X = C_{X,\SM} + C_{X,\NP}$.
One now fits to the data to extract  $C_{X,\NP}$. There are several scenarios that give a good fit to the data and results of 
recent fits can be found in Ref.~\cite{Alok:2019ufo, Ciuchini:2019usw, Aebischer:2019mlg, Alguero:2019pjc,Datta:2019zca, Kowalska:2019ley}.
One of the popular scenario is  $C_{9,\NP}^{\mu\mu} =
-C_{10,\NP}^{\mu\mu}$ which can arise from the  tree-level exchange of leptoquarks (LQ) or a $Z^\prime$ which may be 
heavy \cite{Calibbi:2015kma, Bhattacharya:2016mcc, Greljo:2015mma, Kumar:2018kmr}
or light \cite{Datta:2017pfz, Datta:2017ezo, Alok:2017sui, Datta:2018xty, Sala:2017ihs, Altmannshofer:2017bsz}. Here we will focus on the  LQ solution and there
 are three types of LQ that can generate this scenario.
These are the $SU(2)_L$-triplet scalar ($S_3$), the
$SU(2)_L$-singlet vector ($U_1$), and the $SU(2)_L$-triplet vector
($U_3$). We will focus on the $S_3$ which along with diquarks can be
used to generate neutrino masses at loop level \cite{Kohda:2012sr,Guo:2017gxp}.  
 To generate the neutrino masses, one can  fix the $S_3$ couplings by  a fit to the $\bsll$ data and then the diquark couplings are
 constrained from the neutrino parameters. In this paper we point out that the diquark couplings can be fixed
from nonleptonic $B$ decays and now one can check whether the correct neutrino masses and mixings are reproduced. We would like to mention that 
joint explanation of $R_K^{(*)}$ and $R_D^{(*)}$ was first pointed out in \cite{Bhattacharya:2014wla} and later a connection
between $R_K^{(*)}$ or $R_D^{(*)}$ and neutrino masses was discussed in \cite{Popov:2016fzr,Marzo:2019ldg,Cata:2019wbu}. Here, we are anticipating a common 
framework with leptoquarks and diquarks that can explain the semileptonic and nonleptonic $B$ measurements along with 
the neutrino masses and mixing.

 The observations that we will use for the nonleptonic decays are the set of  $\btopik$ decays. These are penguin dominated nonleptonic $b$ decays and have
 been studied extensively. The decays in the set  include  $B^+ \to \pi^+ K^0$
(designated as $+0$ ), $B^+ \to \pi^0 K^+$ ($0+$), $\bd \to \pi^-
K^+$ ($-+$) and $\bd \to \pi^0 K^0$ ($00$). Their amplitudes are not
independent, but obey a quadrilateral isospin relation:
\beq
\sqrt{2} A^{00} + A^{-+} = \sqrt{2} A^{0+} + A^{+0} ~.
\eeq
Using these decays, nine observables have been measured: the four
branching ratios, the four direct $CP$ asymmetries $A_{CP}$, and the
mixing-induced indirect $CP$ asymmetry $S_{CP}$ in $\bd\to
\pi^0K^0$. Shortly after these measurements were first made (in the
early 2000s), it was noted that there was an inconsistency among
them. This was referred to as the ``$\btopik$ puzzle'' \cite{Buras:2003yc,Buras:2003dj, Buras:2004ub, Baek:2004rp}.
 
 Recently the fits were updated
 \cite{Beaudry:2017gtw, Fleischer:2017vrb, Fleischer:2018bld}.
 In Ref.~\cite{Beaudry:2017gtw}  it was observed that the key input to 
 understanding the data was the ratio of the color-suppressed tree amplitude ($C'$) to the color-allowed
 $(T')$ amplitude.
Theoretically, this ratio is predicted to be $0.15 \lsim
|C'/T'| \lsim 0.5$ \cite{Beneke:2001ev}  with a default value of around 0.2.  It was found that  for a large $|C'/T'| = 0.5$, 
the SM can explain the data satisfactorily.
However, with a  small,
$|C'/T'| = 0.2$, the fit  to the data has a $p$ value of 4\%, which is poor.  Hence,
 if $|C'/T'|$ is small, the SM cannot explain the
$\btopik$ puzzle $-$ NP is needed. The precise statement of the situation is then,
  the measurements of
$\btopik$ decays {\it allow} for NP and  so in this paper we will assume there is NP in these decays.
There are two types of NP mediators that one can consider for the
$\btopik$ decays. 
One is a $Z'$ boson that has a flavor-changing coupling to
${\bar s}b$ and also couples to ${\bar u}u$ and/or ${\bar d}d$. The second
option is a diquark that has $db$ and $ds$ couplings or $ub$ and $us$
couplings.  We will focus on the diquark explanation as the diquarks can contribute to neutrino masses.

The paper is organized in the following manner. In Sec. II we  describe the setup with  leptoquarks and diquarks   that leads to neutrino masses and mixing at the loop level. In that section we also discuss the low energy constraints
for the leptoquark  Yukawa couplings including the $\bsll$ data. 
In Sec. III we explore the $\btopik$ decays mediated by the exchange of diquarks 
and we consider the constraints on the diquark Yukawa couplings from the 
$\btopik$ decays and meson oscillations. In Sec. IV we consider the collider constraints on the  diquark and leptoquarks coupling and masses and we give a scan of all their couplings that satisfy all the constraints and generate the correct neutrino masses and couplings. For a few benchmark cases we present explicit expressions for the diquark and  the leptoquark Yukawa couplings and predict the branching ratios for the rare decays $B \to \phi \pi$ and $B \to \phi \phi$.
Finally in Sec. V we present our conclusions.

\section{Colored Zee Babu Model}
\label{sec:1}
We briefly summarize the main features of the colored Zee Babu model \cite{Babu:2001ex,Kohda:2012sr} that are central to our idea. The model includes a
scalar leptoquark $S_{3L}$(with lepton number 1) of mass $m_L$ and a scalar diquark $S_D$ of mass $m_S$ transforming as
\footnote{The choice $(3,1,-1/3)$ is also possible as it couples neutrinos to down-type quarks but will not explain the $R_K$ and $R^*_K$ anomaly as this scalar
couples up-type quarks to charged leptons.} $(3,3,-1/3)$ and \footnote{Note that if we had chosen the diquark to be $(3,1,-2/3)$, $Y_d$ and, 
hence, the neutrino mass matrix would
be antisymmetric.} $(6,1,-2/3)$ respectively under SM gauge group $SU(3)_c\times SU(2)\times U(1)_Y$ with $Q=T_3+Y$. The baryon number of $S_{3L}$ is taken to be $1/3$ whereas
$S_D$ is assigned $2/3$. With this assignment of baryon number, the baryon conservation is automatic and thus the proton decay is forbidden. The lepton number is 
softly broken through a trilinear term thereby generating Majorana neutrino mass. 

With the particle content discussed above, the interaction Lagrangian is given as
\beq
\mathcal{L}_{int}=- Y_l^{i j}\, \overline{L_i^c}\, i\,\sigma_2\,Q_j^\alpha\, S_{3L}^{\alpha *} 
 - Y_d^{i j}\, \overline{d_{iR}^{\alpha c}}\, d_{jR}^{\,\beta}\,
  S_D^{\alpha\beta *}
 + \mu\, S_{3L}^{\alpha *}\, S_{3L}^{\beta *}\,S_D^{\alpha\beta}+\text(H.c.),
 \label{Lint:1}
\eeq
where $\alpha,\beta=r,b,g$ are $SU(3)_c$ indices, $i,j=1,2,3$ are generation indices, the diquark coupling matrix, $Y_d^{i j}$, is a symmetric complex matrix whereas
the leptoquark coupling matrix, $Y_{l}^{i j}$, is a general complex matrix. The leptoquark couples to leptons and quarks as $\sqrt{2}\nu_{iL}\,u_{jL}-\sqrt{2}
e_{iL}\,d_{jL}+\nu_{iL}\,d_{jL}+e_{iL}\,u_{jl}$. Note that, in Eq. \ref{Lint:1}, we can also have additional scalar interaction terms(not relevant to our
analysis), such as 
\[ \lambda_1\Phi^\dagger\Phi \text{Tr}(S^\dagger_{3L}S_{3L}) + \lambda_2 \text{Tr}(\Phi^\dagger
S_{3L}S^\dagger_{3L}\Phi)\] 
where $\Phi$ is a Higgs doublet. These terms give rise to splitting in the mass of $S_{3L}$ particles, comprising three states of different electric 
charges $-4/3, \,-1/3 \,\, \& \,\, 2/3$, and thus contribute to the oblique corrections\cite{Cheung:2016frv}. To avoid that, 
we assume $\lambda_{1,2}=0$ such that all $S_{3L}$ particles/states have same mass, $m_L$. Along with this, there are quartic and quadratic terms 
of these scalars. We assume that their coefficients are adjusted such that only the Higgs doublet gets the vev and the potential is bounded from 
below.

\begin{figure}[htb]
\centering
\includegraphics{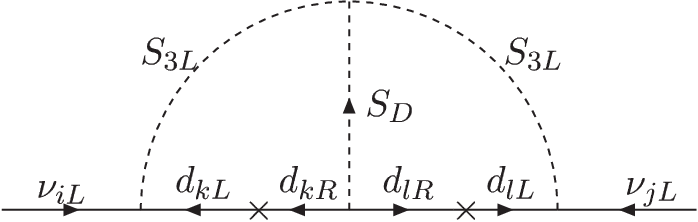} 
\caption{The two loop neutrino mass generated by $(3,3,-1/3)$ leptoquark and $(6,1,-2/3)$ diquark.}
\label{diag:nmass}
\end{figure}

The above Lagrangian can generate majorana neutrino mass at two loop as depicted in the Fig. \ref{diag:nmass}. The resultant neutrino 
matrix is given as~\cite{AristizabalSierra:2006gb,Kohda:2012sr}
\begin{equation}
M_{\nu}^{i\,j} = 24\, \mu \,Y_l^{ik} m_d^{kl}\, Y_{d}^{lm} I^{lm} m_d^{mn}\, Y_l^{nj},
\label{masnu}
\end{equation}
where $I^{kl}$ is a loop integral, which in the limit of large leptoquark and diquark masses simplifies to
\begin{equation}
I^{kl}\simeq \dis
\frac{1}{(4 \pi)^4} \frac{1}{m_L^2} \tilde{I}\left(\frac{m_{S}^2}{m_L^2}\right),
\end{equation}
with
\begin{equation}
\tilde{I}(r)=\int_0^1dx\int_0^{1-x}dy \frac{1}{x+y(y+r-1)}
\ln\left(\frac{x+ry}{y(1-y)}\right),
\end{equation}
and $m_d$ is 3$\times$3 diagonal mass matrix for down-type quarks. Note that we have chosen diagonal 
bases of the mass matrix for down-type quarks and charged leptons. Hence, to obtain the correct masses 
of neutrino, we need to diagonalize the mass matrix, $M_\nu$ by the PMNS matrix $\mathcal{U}$ as
\begin{equation}
m_\nu = \mathcal{U}^\dagger M_\nu \mathcal{U}.
\end{equation}
The standard parametrization is adopted such that
\beq
\label{eq:definition_of_pmns}
 \mathcal{U}=\left(\begin{array}{ccc} 1 & 0 & 0\\
0 & c_{23} & s_{23}\\
0 & -s_{23} & c_{23}\end{array}\right)\left(\begin{array}{ccc}
c_{13} & 0 & s_{13} e^{-i\delta} \\
0 & 1 & 0 \\
-s_{13} e^{i\delta} & 0 & c_{13}\end{array}
\right)\left(\begin{array}{ccc}
c_{12} & s_{12} & 0 \\
-s_{12} & c_{12} & 0 \\
0 & 0 & 1\end{array}\right)
\begin{pmatrix}
1 & 0 & 0
\\
0 & e^{  i \alpha_{21}/2} & 0
\\
0 & 0 & e^{  i \alpha_{31}/2}
\end{pmatrix}
\eeq
where $c_{ij}$ and $s_{ij}$ represent $\cos\theta_{ij}$ and $\sin\theta_{ij}$, respectively. In the case of Majorana neutrinos,
 $\alpha_{21}$ and $\alpha_{31}$ are the extra $CP$ phases that cannot be determined by the oscillation experiments. However, these phases
 could be sensitive to the upcoming neutrinoless double beta decay searches. 
 
It should be noted that the mass dimension one parameter, $\mu$, is constrained by demanding the perturbativity of the theory.
The trilinear term in the Eq. \ref{Lint:1} generates one-loop corrections to leptoquark and diquark masses. These corrections($\Delta
m^2$) are, in general, proportional to $\dis \frac{\mu^2}{16\pi^2}$. Requiring corrections to be smaller than the corresponding masses
implies $\dis \mu \ll \dis 4\pi m_{S/L}$\cite{AristizabalSierra:2006gb}. As various collider searches, discussed in Sec. \ref{sec:4},
do not allow the scalar masses to be smaller than 1 TeV, we take $\mu$ from 0.1$-$1 TeV and this choice commensurates
with the above constraints.

Having discussed the details of the model, next$-$we list all the possible constraints, coming from various experiments on leptoquark and
diquark coupling matrices.

\section{Leptoquarks}
\label{sec:2}
\begin{itemize} 
 \item \textbf{Lepton flavour violation at tree level}: Collider searches of leptoquarks indicate that they are heavy. So we can study 
 their low energy effects by writing 4-Fermi operators of two lepton-two quarks. Using Fierz rearrangement, we get
 
 \[\frac{Y_l^{ik} Y_l^{jn*}}{2 m^2_L}(\overline{l_i}\gamma^\mu P_L l_j)(\overline{q_k} \gamma_\mu P_L q_n) + \text{H.c}\]
 
  as an effective operator where $l$ and $q$ denote leptons and quarks. These are organized in terms of the four-Fermi effective interactions 
 with normalized dimensionless Wilson coefficients as
 \[\mathcal{H}_{\text{eff}}=\sum_{ijkn}\frac{Y_l^{ik} Y_l^{jn*}}{2 m^2_L}\mathcal{O}_{ijkn}=\frac{-4G_F}{\sqrt{2}}\sum_{ijkn}C^{ijkn}\mathcal{O}_{ijkn}\].
 In Ref.\cite{Carpentier:2010ue}, constraints on such operators have been extensively studied. Keeping in mind that $Y_l^{ij}$ should be 
 able to explain a small neutrino mass, following are the most crucial operators related to our work:
 \begin{itemize}
  \item $(\overline{e_i}\gamma^\mu P_L e_j)(\overline{d} \gamma_\mu P_L d)$: The $\mu-e$ conversion in nuclei sets a bound on the
  Wilson coefficient of this operator, i.e.
  \begin{eqnarray}
  C^{1211}
  =\left|\frac{Y_l^{11} Y_l^{21*}}{4\sqrt{2} G_F m_L^2}\right| < 8.5\times 10^{-7}.
  \end{eqnarray}
  \item $(\overline{\mu}\gamma^\mu P_L e)(\overline{d} \gamma_\mu P_L s)$: The bound from the decay $K^\circ\to e^+\mu^-$ sets a 
  bound on $C^{1212}$
	\begin{eqnarray}
	C^{1212}
	=\left|\frac{Y_l^{12}Y_l^{21*}}{4\sqrt{2}G_F \,m_L^2}\right|< 3.0\times 10^{-7}.
	\end{eqnarray}
  \item $(\overline{\nu_i}\gamma^\mu P_L \nu_j)(\overline{d_k} \gamma_\mu P_L d_l)$: The constraint on the $K$ meson decay to pion and 
  neutrinos($\nu_i\nu_j$)
  sets another bound:
  \begin{eqnarray}
  C^{ij 12}
  =\left|\frac{Y_l^{i1}Y_l^{j2*}}{4\sqrt{2}G_F\,m_L^2}\right|<9.4\times 10^{-6},
  \end{eqnarray}
 \end{itemize}
 Apart from this, we have also taken care of all the relevant Wilson coefficients mentioned in Ref.\cite{Carpentier:2010ue}.
 \item \textbf{Lepton flavour violation radiative decay}:
 The LFV radiative decays $l_i \to l_j \gamma$ are induced at one loop by the exchange 
 of a leptoquark $S_{3L}$  with the branching ratio~\cite{Cheung:2016frv}
  \begin{equation}
  \text{BR}(\ell_i\to\ell_j \gamma) \simeq \frac{{3}\,\alpha\, \chi_i}{256 \pi G_F^2}\frac{1}{m_L^4}
  \big|(Y_lY_l^\dagger)^{ij}\big|^2
  \end{equation}
  where $\alpha = \dis\frac{e^2}{4\pi}$, $\chi_\mu=1$, and $\chi_\tau=1/5$. In the case of a $\tau$ lepton, there are two leptonic modes and 
  hadronic modes can be approximated by a single partonic mode(with three colors). Hence there is a factor of 5 difference in $\mu$
 and the $\tau$-lepton branching ratio.  
 The current experimental bounds\cite{TheMEG:2016wtm,Aubert:2009ag} are 
  \begin{itemize}
  \item $\text{BR}(\mu\to e\gamma)<4.2\times10^{-13}$, 
  \item $\text{BR}(\tau\to \mu\gamma)<4.4\times10^{-8}$,
  \item $\text{BR}(\tau\to e\gamma)<3.3\times10^{-8}$.
 \end{itemize}
 
  \item \textbf{$\mathbf{\bsll}$ anomalies}:
As discussed in the Introduction one can perform fits to the $\bsll$ data and scenarios in terms of Wilson's coefficients
that give a good description of the data. In the above set up, the exchange of  the $S_{3L}$ leptoquark at tree level
contributes to the decay $b \to s \ell^+ \ell^-$,  and in particular generates the scenario   $C_{9,\NP}^{\mu\mu} =
-C_{10,\NP}^{\mu\mu}$.
The effective Hamiltonian describing the decay is parameterized as
\begin{equation}
	\mathcal{H}_{\text{eff}} = - \frac{4 G_F}{\sqrt{2}} \frac{\alpha}{4 \pi} V_{tb} V_{ts*} \sum_i C_i(\mu) \mathcal{O}_i(\mu) + \text{H.c.},
\end{equation}
where $\mathcal{O}_i(\mu)$ are effective operators with Wilson coefficients $C_i(\mu)$ renormalized at the scale $\mu$. 
For the model under consideration, only the operators $\mathcal{O}_9^{\ell_i}=(\bar{s}\gamma^\mu P_L b)(\bar{\ell}_i\gamma^\mu \ell_i)$ and 
$\mathcal{O}_{10}^{\ell_i}=(\bar{s}\gamma^\mu P_L b)(\bar{\ell}_i\gamma^\mu\gamma_5 \ell_i)$ are induced. Using Fierz identity, we obtain  
the following Wilson coefficients:
\begin{equation}
C_9^{\ell_i} = -C_{10}^{\ell_i} =
-\frac{\sqrt{2}\pi}{4\alpha G_Fm_{L}^2}
\frac{(Y_l^{ i3})(Y_l^{i2*})}{V_{tb}V_{ts*}}.
\end{equation}
Assuming new physics only in the muon sector, a model independent analysis on the above operators \cite{Datta:2019zca} from the $R_K$, $R^*_K$, $P'_5$  and other observables suggests that 
\[C_9^{\mu\mu}(\text{NP}) = -0.53 \pm 0.08. \] 
\end{itemize}

\section{Diquark}
\label{sec:3}
\subsection{\boldmath Nonleptonic decays and the $\btopik$ puzzle}
In the Standard Model (SM) 
the amplitudes for hadronic $B$ decays of the type $b\to q \bar{f} f$ 
are generated by the following effective 
Hamiltonian:
\begin{eqnarray}
H_{eff}^q &=& {G_F \over \protect \sqrt{2}} 
   [V_{fb}V^*_{fq}(c_1O_{1f}^q + c_2 O_{2f}^q) -
     \sum_{i=3}^{10}
V_{tb}V^*_{tq} c_i^t O_i^q] +H.c.\;,
\end{eqnarray}
where the
superscript  $t$
indicates the internal quark, and $f$ can be a $u$ or 
$c$ quark. $q$ can be either a $d$ or an $s$ quark depending on 
whether the decay is a $\Delta S = 0$
or $\Delta S = -1$ process.
The operators $O_i^q$ are defined as
\begin{eqnarray}
O_{f1}^q &=& \bar q_\alpha \gamma_\mu Lf_\beta\bar
f_\beta\gamma^\mu Lb_\alpha\;,\;\;\;\;\;\;O_{2f}^q =\bar q
\gamma_\mu L f\bar
f\gamma^\mu L b\;,\nonumber\\
O_{3,5}^q &=&\bar q \gamma_\mu L b
\bar q' \gamma^\mu L(R) q'\;,\;\;\;\;\;\;\;O_{4,6}^q = \bar q_\alpha
\gamma_\mu Lb_\beta
\bar q'_\beta \gamma^\mu L(R) q'_\alpha\;,\\
O_{7,9}^q &=& {3\over 2}\bar q \gamma_\mu L b  e_{q'}\bar q'
\gamma^\mu R(L)q'\;,\;O_{8,10}^q = {3\over 2}\bar q_\alpha
\gamma_\mu L b_\beta
e_{q'}\bar q'_\beta \gamma^\mu R(L) q'_\alpha\;,\nonumber
\end{eqnarray}
where $R(L) = 1 \pm \gamma_5$, 
and $q'$ is summed over $u, d, s, c and b$.  $O_2$ and $O_1$ are the tree-level
and QCD corrected operators, respectively. 
$O_{3-6}$ are the strong gluon induced
penguin operators, and operators 
$O_{7-10}$ are due to $\gamma$ and $Z$ exchange (electroweak penguins) 
and ``box'' diagrams at loop level. The Wilson coefficients
 $c_i^f$ are defined at the scale $\mu \approx m_b$ 
and have been evaluated to next-to-leading order in QCD.
The $c^t_i$ are the regularization scheme 
independent values and can be found in Ref. \cite{Beneke:2001ev}.

 The diquarks  discussed in Sec. II in the context of  neutrino mass generation can contribute to the $\btopik$ decays
 and  we can write down the new physics operators  that will be generated  by a 6 or $\overline{3}$  diquark \cite{Giudice:2011ak}.
In the general case we get the effective Hamiltonian for $b$ quark decays $ b \to \bar{d}_i d_j d_k $ as
\bea
{\cal H}^d_{\sss NP} & = & X^d  \, {\bar d}_{\alpha, k} \gamma_\mu( 1 +\gamma^5) b_\alpha 
\, {\bar d}_{\beta, j} \gamma^\mu(1+ \gamma^5) d_{\beta, i},  \
\label{Heff}
\eea
where the superscript $d$ in $X^d$ equals  6 or $\overline{3}$ corresponding to the color sextet or the antitriplet  diquark.
The greek subscripts represent color and the latin subscripts the flavor.
 We have
\bea
X^d& = & -\frac{ Y_{i3}^dY_{jk}^{*d}}{4m_S^2},  \
\eea
where the Yukawa $Y$ are symmetric for the sextet diquark  and antisymmetric for the antitriplet diquark and we have assumed the same masses for the diquarks. 

For $b$ decays of the type $ b \to \bar{s} ss$ the diquark contribution is tiny as the effective Hamiltonian is proportional to $Y_{22}^d$ which vanishes for  the $\overline{3}$ diquark and is highly suppressed from $K$ and $B$ mixing  for the sextet diquark. Similarly
the $ b \to \bar{d} dd$ transition is proportional to $Y_{11}^d$, which is also small.

{} For $ b  \to s \bar{d} d$( $ b \to \bar{d} s d$ and $ b \to \bar{d} ds$) transitions we have the following Hamiltonian:
\bea
{\cal H}^d_{\sss NP} & = & X^d \, {\bar s}_\alpha \gamma_\mu( 1 +\gamma^5) b_\alpha 
\, {\bar d}_{\beta} \gamma^\mu(1+ \gamma^5) d_{\beta}  +
X^{d}_{C}  \, {\bar s}_\alpha \gamma_\mu( 1 + \gamma^5) b_\beta  
\, {\bar d}_{\beta} \gamma^\mu(1+ \gamma^5) d_{\alpha} , \
\label{setupI}
\eea
with
\bea
X^d & = & -\frac{ Y_{13}^dY_{12}^{*d}}{4m_S^2}, \nonumber\\
X^{d}_C & = & -\frac{ Y_{13}^dY_{21}^{*d}}{4m_S^2}, \
\eea
and
\bea
X^{ \overline{3}} &= &-X^{\overline {3}}_{C} ,\nonumber\\
X^{ 6}&= &X^{6}_{C}. \
\label{WC}
\eea

We can rewrite the effective Hamiltonian after a color Fierz transformation as
\bea
{\cal H}^d_{\sss NPF} & = & X^i  \, {\bar d}_\beta \gamma_\mu( 1 + \gamma^5) b_\alpha 
\, {\bar s}_{\alpha} \gamma^\mu(1+ \gamma^5) d_{\beta}  +
X^{i}_{C}  \, {\bar d}_\beta \gamma_\mu( 1 + \gamma^5) b_\beta  
\, {\bar s}_{\alpha} \gamma^\mu(1- \gamma^5) d_{\alpha} . \
\label{setupIF}
\eea

The only other unsuppressed transition is   $ b  \to s \bar{s} d$( $ b \to \bar{s} s d$ and $ b \to \bar{s} ds$)  which has the effective Hamiltonian,
\bea
{\cal H}^d_{\sss NP} & = & X^d \, {\bar s}_\alpha \gamma_\mu( 1 +\gamma^5) b_\alpha 
\, {\bar d}_{\beta} \gamma^\mu(1+ \gamma^5) s_{\beta}  +
X^{d}_{C}  \, {\bar s}_\alpha \gamma_\mu( 1 + \gamma^5) b_\beta  
\, {\bar d}_{\beta} \gamma^\mu(1+ \gamma^5) s_{\alpha} , \
\label{setupII}
\eea
with
\bea
X^d & = & -\frac{ Y_{23}^dY_{12}^{*d}}{4m_S^2}, \nonumber\\
X^{d}_C & = & -\frac{ Y_{23}^dY_{21}^{*d}}{4m_S^2}, \
\eea
In this case at the meson level we can have the decays $B \to \phi \pi$ and the annihilation decays
$B \to \phi \phi$. These decays are highly suppressed in the SM and the observance of these decays could signal the presence of diquarks

\subsection{\bf \boldmath Naive $\btopik$ puzzle}
We begin by reviewing the $\btopik$ puzzle. As in Ref.~\cite{Beaudry:2017gtw} we can analyze the $\btopik$ decays in terms of topological
amplitudes.
Including only the leading diagrams the $\btopik$ amplitudes become
\bea
\label{reducedamps}
A^{+0} &=& -P'_{tc} ~, \nn\\
\sqrt{2} A^{0+} &=& -T' e^{i\gamma} + P'_{tc} - \pewp ~, \nn\\
A^{-+} &=& -T' e^{i\gamma} + P'_{tc} ~, \nn\\
\sqrt{2} A^{00} &=& - P'_{tc} - \pewp ~.
\eea
Here, $T'$ is the color-allowed tree amplitude,  $P'_{tc}$ is the gluonic penguin amplitude, and
 $\pewp$  is the color-allowed electroweak penguin amplitude. Furthermore in the SU(3) limit the $T'$ and 
  $\pewp$ are proportional to each other and so have the same strong phases.
 Now consider the direct $CP$ asymmetries of $B^+ \to
\pi^0 K^+$ and $\bd \to \pi^- K^+$. Such $CP$ asymmetries are generated
by the interference of two amplitudes with nonzero relative weak and
strong phases. In both $A^{0+}$ and $A^{-+}$, $T'$-$P'_{tc}$
interference leads to a direct CP asymmetry. On the other hand, in
$A^{0+}$, $\pewp$ and $T'$ have the same strong phase, $\pewp \propto
T'$ , while $\pewp$ and $P'_{tc}$ have the
same weak phase ($=0$), so that $\pewp$ does not contribute to the
direct $CP$ asymmetry. This means that we expect $A_{CP}(B^+ \to \pi^0
K^+) = A_{CP}(\bd \to \pi^- K^+)$.

The latest $\btopik$ measurements are shown in Table \ref{tab:data}.
Not only are $A_{CP}(B^+ \to \pi^0 K^+)$ and $A_{CP}(\bd \to \pi^-
K^+)$ not equal, they are of opposite sign!  Experimentally, we have
$(\Delta A_{CP})_{\rm exp} = (12.2 \pm 2.2) \%$. This differs from 0
by $5.5\sigma$.  This is the naive $\btopik$ puzzle.

\begin{table}[tbh]
\center
\begin{tabular}{|c|c|c|c|}
\hline
Mode & $BR[10^{-6}]$ & $A_{CP}$ & $S_{CP}$ \\ \hline
$B^+ \to \pi^+ K^0$ & $23.79 \pm 0.75$ & $-0.017 \pm 0.016$ & \\
\hline
$B^+ \to \pi^0 K^+$ & $12.94 \pm 0.52$ & $0.040 \pm 0.021$ & \\
\hline
$\bd \to \pi^- K^+$ & $19.57 \pm 0.53$ & $-0.082 \pm 0.006$ & \\
\hline
$\bd \to \pi^0 K^0$ & $9.93 \pm 0.49$ & $-0.01 \pm 0.10$ & $0.57 \pm 0.17$ \\
\hline
\end{tabular}
\caption{Branching ratios, direct $CP$ asymmetries $A_{CP}$, and
  mixing-induced $CP$ asymmetry $S_{CP}$ (if applicable) for the four
  $\btopik$ decay modes. The data are taken from Ref.~\cite{HFAGWinter2016}.}
\label{tab:data}
\end{table}

\subsection{\bf Model-independent new physics formalism}
\label{ModelindepNP}
In the general approach of Refs.~\cite{Datta:2004re,Datta:2004jm}, the NP operators that
contribute to the $\btopik$ amplitudes take the form ${\cal
  O}_{\rm NP}^{ij,q} \sim {\bar s} \Gamma_i b \, {\bar q} \Gamma_j q$ ($q
= u,d$), where $\Gamma_{i,j}$ represents Lorentz structures, and color
indices are suppressed. The NP contributions to $\btopik$ are encoded
in the matrix elements $\bra{\pi K} {\cal O}_{\rm NP}^{ij,q} \ket{B}$. In
general, each matrix element has its own NP weak and strong phases.

Note that the strong phases are basically generated by QCD
rescattering from diagrams with the same CKM matrix elements. One can
argue that the strong phase of $T'$ is expected to be very small
since it is due to self-rescattering. For the same reason, all NP
strong phases are also small, and can be neglected. In this case, many
NP matrix elements can be combined into a single NP amplitude, with a
single weak phase:
\beq
\sum \bra{\pi K} {\cal O}_{NP}^{ij,q} \ket{B} = \ANPq
e^{i \Phi_q} ~.
\eeq
Here the strong phase is zero. There are two classes of such NP
amplitudes, differing only in their color structure: ${\bar s}_\alpha
\Gamma_i b_\alpha \, {\bar q}_\beta \Gamma_j q_\beta$ and ${\bar
  s}_\alpha \Gamma_i b_\beta \, {\bar q}_\beta \Gamma_j q_\alpha$ ($q
= u,d$). They are denoted $\ApNPqph$ and $\ApNPCqph$, respectively
\cite{Datta:2004jm}. Here, $\Phi'_q$ and $\Phi_q^{\prime {C}}$ are the
NP weak phases. In general, ${\cal A}^{\prime,q} \ne {\cal A}^{\prime
  {C}, q}$ and $\Phi'_q \ne \Phi_q^{\prime {C}}$. Note that, despite
the ``color-suppressed'' index $C$, the matrix elements $\ApNPCqph$
are not necessarily smaller than $\ApNPqph$.

There are therefore four NP matrix elements that contribute to
$\btopik$ decays. However, only three combinations appear in the
amplitudes: $\ApNPcomb \equiv - \ApNPuph + \ApNPdph$, $\ApNPCuph$, and
$\ApNPCdph$ \cite{Datta:2004jm}. The $\btopik$ amplitudes can now be
written in terms of the SM diagrams and these NP matrix elements. Here
we neglect the small SM diagram $P'_{uc}$ but include the color-suppressed amplitudes:
\bea
\label{BpiKNPamps}
A^{+0} &=& -P'_{tc} -\frac13 \pewcp + \ApNPCdph ~, \nn\\
\sqrt{2} A^{0+} &=& P'_{tc} - T' \, e^{i\gamma} - \pewp -C' \, e^{i\gamma} -\frac23 \pewcp + \ApNPcomb - \ApNPCuph ~, \nn\\
A^{-+} &=& P'_{tc} - T' \, e^{i\gamma} -\frac23 \pewcp - \ApNPCuph ~, \nn\\
\sqrt{2} A^{00} &=& -P'_{tc} - \pewp -C' \, e^{i\gamma} -\frac13 \pewcp + \ApNPcomb + \ApNPCdph ~.
\eea

 We can express the various matrix elements as
\bea
 \ApNPCdph  & = &  \sqrt{2} \bra{ \pi^0 K^0}{\cal H}^d_{\sss NPF}\ket{B^0} =   \bra{ \pi^+ K^0}{\cal H}^d_{\sss NPF}\ket{B^+},\nonumber\\
 \ApNPCuph    & = &- \sqrt{2} \bra{ \pi^0 K^+}{\cal H}^u_{\sss NPF}\ket{B^+} =    \bra{ \pi^- K^+}{\cal H}^u_{\sss NPF}\ket{B^0},\nonumber\\ 
 \ApNPcomb & = &  \sqrt{2} \bra{ \pi^0 K^+}\left[ {\cal H}^u_{\sss NP} +{ \cal H}^d_{\sss NP} \right]\ket{B^+}    =   \sqrt{2} \bra{ \pi^0 K^0} \left[{\cal H}^u_{\sss NP} +{ \cal H}^d_{\sss NP} \right]\ket{B^0}. \
 \label{MEdef}
 \eea
 In our model
 ${\cal H}^u_{\rm NP} $ and  ${\cal H}^u_{\rm NPF}$ are absent while
  ${\cal H}^d_{\rm NP} $ and  ${\cal H}^d_{\rm NPF}$  are defined in Eqs.~\ref{setupI}  and ~\ref{setupIF}.
 In the factorization assumption and using Eqs.~\ref{setupI}  and ~\ref{setupIF} we get the following results
 for the nonzero amplitudes,
 \bea
 \ApNPCdph  & = &\left[ X^{6} - X^{\overline {3}}  +\frac{X^{6} + X^{\overline {3}}}{N_c} \right]   \bra{ \pi^+} \, {\bar d}_\beta \gamma_\mu( 1 + \gamma^5) b_\beta  \ket{B^+}
\bra{K^0}\, {\bar s}_{\alpha} \gamma^\mu(1+ \gamma^5) d_{\alpha} \ket{0}, \nonumber\\
\ApNPdph & = & \sqrt{2}  \left[ X^{6} +X^{\overline {3}}  +\frac{X^{6} - X^{\overline {3}}}{N_c} \right]   \bra{ K^+} \, {\bar s}_\beta \gamma_\mu( 1 + \gamma^5) b_\beta  \ket{B^+}
\bra{\pi^0}\, {\bar d}_{\alpha} \gamma^\mu(1+ \gamma^5) d_{\alpha} \ket{0} . \
\label{setupI_me}
\eea
In Ref.~\cite{Baek:2009pa}, a different set of NP operators is defined:
\bea
\pewpnp \, e^{i \Phi'_{\rm EW}} & \equiv & \ApNPuph - \ApNPdph ~, \nn\\
P'_{NP} \, e^{i \Phi'_{P}} & \equiv & \frac13 \ApNPCuph\ + \frac23 \ApNPCdph~, \nn\\
\pewcpnp \, e^{i \Phi^{\prime {C}}_{\rm EW}} & \equiv & \ApNPCuph\ - \ApNPCdph ~.
\label{NPPoperators}
\eea
In this case we have
\bea
\pewpnp \, e^{i \Phi'_{EW}} & \equiv &  - \ApNPdph ~, \nn\\
P'_{NP} \, e^{i \Phi'_{P}} & \equiv &  \frac23 \ApNPCdph~=  -(2/3) \pewcpnp  \nn\\
\pewcpnp \, e^{i \Phi^{\prime {C}}_{EW}} & \equiv & - \ApNPCdph ~.
\label{NPPoperators}
\eea
\begin{table}[H]
\center
\begin{tabular}{|c|c|}
\hline
\multicolumn{2}{|c|}{NP fit (1): $\chi^2/{\rm d.o.f.} = 3.75/4$,}\\
\multicolumn{2}{|c|}{~~~~~~~~~~~~~~~~~~ p-value $=0.44$}\\
 \hline
Parameter & Best-fit value \\ 
\hline
$\gamma$ & $(67.5 \pm 3.4)^\circ$ \\
\hline
$\beta$ & $(21.80 \pm 0.68)^\circ$ \\
\hline
$\Phi'$ & $(37.0 \pm 12.6)^\circ$ \\
\hline
$|T'|$ & $19.1 \pm 2.8$ \\
\hline
$|P'_{tc}|$ & $48.7 \pm 1.2$ \\
\hline
$\pewpnp$ & $8.6 \pm 2.5$ \\
\hline
$\pewcpnp$ & $2.7 \pm 1.1$ \\
\hline
$\delta_{P'_{tc}}$ & $(-4.0 \pm 1.1)^\circ$ \\
\hline
$\delta_{C'}$ & $(-60.0 \pm 115.6)^\circ$ \\
\hline
\end{tabular} 
\qquad \quad
\begin{tabular}{|c|c|}
\hline
\multicolumn{2}{|c|}{NP fit (2): $\chi^2/{\rm d.o.f.} = 3.82/4$,}\\
\multicolumn{2}{|c|}{~~~~~~~~~~~~~~~~~~ p-value $=0.43$}\\
 \hline
Parameter & Best-fit value \\ 
\hline
$\gamma$ & $(74.7 \pm 5.2)^\circ$ \\
\hline
$\beta$ & $(21.80 \pm 0.68)^\circ$ \\
\hline
$\Phi'$ & $(18.7 \pm 33.9)^\circ$ \\
\hline
$|T'|$ & $19.7 \pm 7.1$ \\
\hline
$|P'_{tc}|$ & $45.5 \pm 3.9$ \\
\hline
$\pewpnp$ & $6.7 \pm 3.9$ \\
\hline
$\pewcpnp$ & $6.5 \pm 3.7$ \\
\hline
$\delta_{P'_{tc}}$ & $(-4.0 \pm 2.0)^\circ$ \\
\hline
$\delta_{C'}$ & $(-48.9 \pm 23.5)^\circ$ \\
\hline
\end{tabular}
\caption{$\chi^2_{\rm min}/{\rm d.o.f.}$ and best-fit values of
  unknown parameters for the Diquark model where the fit 1 has $X^6=X^{\overline{3}}$, and fit 2  has $X^{\overline{3}}=0$. 
  Constraints: $\btopik$ data, measurements of $\beta$ and
  $\gamma$, $|C'/T'| = 0.2$, $|\pewcpnp/\pewpnp| = 0.3$ (fit 1), and $|\pewcpnp/\pewpnp|=1$ (fit 2).}
\label{tab:NPfitcase3}
\end{table}

We consider two models, the first with
\bea
 X^6 & = & X^{\overline {3}} \
 \eea
This  leads to  $\pewcpnp/\pewpnp = { 1 \over 3}$ with both amplitudes having the same weak phase. 
\bea
\pewpnp \, e^{i \Phi'_{EW}} & \equiv & 
\frac{ Y_{d13}^6Y_{d12}^{*6}}{4m_S^2} 
   \sqrt{2}     \bra{ K^+} \, {\bar s}_\beta \gamma_\mu( 1 + \gamma^5) b_\beta  \ket{B^+} 
 \bra{0}\, {\bar s}_{\alpha} \gamma^\mu(1+ \gamma^5) d_{\alpha} \ket{K^0}   
 ~, \nn\\
P'_{NP} \, e^{i \Phi'_{P}} & \equiv &  \frac23 \ApNPCdph~=  -(2/3) \pewcpnp  \nn\\
\pewcpnp \, e^{i \Phi^{\prime {C}}_{EW}} & \equiv & - \ApNPCdph=  \pewpnp \, e^{i \Phi'_{EW}} /3  ~.
\label{NPPoperators}
\eea
The second model has 
\bea
X^{\overline{3}} & = & 0. \
\eea
This leads to $\pewcpnp/\pewpnp = 1$, again with both amplitudes having the same weak phase.

A $\chi^2$ fit for the new physics within this scenario is
performed to determine the parameters of the model. The procedure for determining such a fit is as follows. We define the function
\bea
\label{chisquare}
\chi^2=\sum_{i=1}^N\ \left(\frac{\mathcal{O}_{\rm exp}-\mathcal{O}_{\rm th}}{\Delta \mathcal{O}_{\rm exp}}\right)^2
\eea
where $\mathcal{O}_{\rm exp}$ and $\Delta\mathcal{O}_{\rm exp}$ are the experimentally determined quantities with
their associated uncertainties, respectively, as listed in Table \ref{tab:data}. $\mathcal{O}_{th}$ are determined
from the model and are thus functions of the unknown parameters. The goal from here is to find the values of the 
parameters that minimize $\chi^2$. There are many programs available to accomplish this, one of the most widely used
is {\small MINUIT}~\cite{James:1994vla}, which is used here. The goodness of the fit is determined by the value of $\chi^2$ at
the minimum and the number of degrees of freedom in the fit. The degrees of freedom are the number of constraints 
included in the fit minus the number of parameters that are fitted. In this case the number of constraints is 13: 
the $B\to \pi K$ data, the independent measurements of $\beta$ and $\gamma$, and the constraints on 
$|C^\prime/T^\prime|$ and $|P_{EW,NP}^{\prime C} / P_{EW,NP}^\prime|$. The number of parameters is nine and we have 
that the number of degrees of freedom are four. A ``good'' fit is one where $\chi^2_{\rm min}\approx$ d.o.f., but a better
measure is the $p$ value which gives the probability that the model tested adequately describes the observations. 

The results of the fit for
this case are shown in Table \ref{tab:NPfitcase3}. Here the $p$ value is
44\% for $X^6=X^{\overline{3}}$, and 43\% for $X^{\overline{3}}=0$, which is not bad (and is far better than that of the SM).
%

%

\begin{figure*}[htb]
\centering
 \includegraphics[width=75mm]{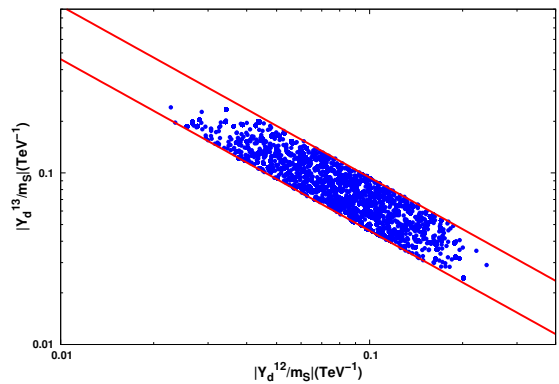}
\caption{\em The correlation between $\dis\frac{|Y_d^{12}|}{m_{S}}$ and $\dis\frac{|Y_d^{13}|}{m_{S}}$ within $1\sigma$ range. The shaded area corresponds 
to mass range $m_S \in [5:20]$\tev.}
\label{Fig:di2} 
\end{figure*}

The SM $T'$ diagram involves the
tree-level decay ${\bar b} \to {\bar u} W^{+*} (\to u {\bar s} =
K^+)$.  The NP $\pewpnp$ diagram looks very similar
and is expressed relative to the $T'$ diagram.
 Within factorization, the SM and NP diagrams involve
$A_{\pi K} \equiv F_0^{B \to \pi}(0) f_K$ and $A_{K \pi} \equiv F_0^{B
  \to K}(0) f_\pi$, respectively, where $F_0^{B \to K,\pi}(0)$ are
form factors and $f_{\pi,K}$ are decay constants. The hadronic factors
are similar in size: $|A_{K \pi}/A_{\pi K}| = 0.9 \pm 0.1$
\cite{Beneke:2001ev}.  Taking central values for $X^6=X^{\overline{3}}$, we have \cite{Beaudry:2017gtw}
%
%
\bea
\label{constraint1}
&&\Phi'= Arg[Y_{d13}^{{6}}Y_{d12}^{*{6}}] \nonumber\\
&& \left\vert \frac{\pewpnp}{T'} \right\vert \simeq 
\frac{2 A_{K \pi}  |X^{\overline {3}}|}  
{A_{\pi K} (G_F/\sqrt{2})|V_{ub*} V_{us}|} = \frac{8.6}{19.1} \nn\\
&\Longrightarrow&  \left\vert \frac{ Y_{d13}^{6}Y_{d12}^{*6}}{2m_S^2} \right \vert   = (3.4 \pm 1.2) \times 10^{-3} ~ {\rm TeV}^{-2} ~.
\eea
For $X^{\overline{3}}=0$ we obtain
\bea
\label{constraint2}
\left|\frac{Y^6_{d13}Y^{*6}_{d12}}{2m_S^2}\right|=(2.6 \pm 1.8) \times 10^{-3} ~ {\rm TeV}^{-2}
\eea
Both models give similar fits and in Fig.~\ref{Fig:di2}  we show the allowed regions of the diquark couplings
within a 1$ \sigma$ range for the first model.

\subsection{Neutral meson Mixing}
Diquarks, in spite of being charged, through their coupling to the same generation quarks can
mediate the mixing between neutral mesons at tree level. Following the convention in \cite{Bona:2007vi}, the 
mixing can be depicted as the six dimension operator:
\[ \mathcal{O}_{mix} = \frac{Y^{*ij}_d Y_d^{kl}}{m^2_{S}}\bar{\psi^k}_R\gamma^\mu {\psi^i}_R \, \bar{\psi^l}_R\gamma_\mu {\psi^j}_R\]
The 90 \% C.L. bounds on the corresponding Wilson coefficients\cite{Bona:2007vi} is then given as
\begin{equation*}
 \begin{array}{rcrcl}
 \mathbf{K^\circ-\overline{K^\circ}} &&	 \qquad \quad           \displaystyle\left|\frac{Y^{*11}_d Y_d^{22}}{4\sqrt{2}G_F m_S^2}\right|& <&  2.9\times 10^{-8} \\
 \mathbf{B_d^\circ-\overline{B_d^\circ}} &&  \qquad \quad	\displaystyle\left|\frac{Y^{*11}_d Y_d^{33}}{4\sqrt{2}G_F m_S^2}\right|& <& 7.0\times 10^{-7} \\
 \mathbf{B_s^\circ-\overline{B_s^\circ}} && \qquad \quad	\displaystyle\left|\frac{Y^{*22}_d Y_d^{33}}{4\sqrt{2}G_F m_S^2}\right|& <&  3.3\times 10^{-5} 
 \end{array}
\end{equation*}

\section{Numerical Analysis and Discussion}
\label{sec:4}
 Before we present the results, we discuss the bounds on the scalar masses obtained from collider experiments. The collider 
 experiments provide direct limits on the leptoquark mass when they decay to leptons and quarks in the final state. There are many
 studies in the literature where different signatures have been discussed\cite{Bhattacharyya:1994yc,Bandyopadhyay:2018syt,Kohda:2012sr}. 
 The leptoquarks can be pair produced from $gg$ and $q\bar{q}$ as initial state or singly produced at hadron colliders via $g \, +\,q\rightarrow
 S_{3L} + \text{lepton}$. Recent studies at ATLAS\cite{Aaboud:2016qeg} and CMS\cite{Sirunyan:2018nkj} with 13 TeV data puts a bound on the
 scalar leptoquark mass, $m_{L} > 1,1.2\,\,(\text{ATLAS}),0.9\,\,(\text{CMS})\,\, \tev$ when decay to $u\, e$, $c\, \mu$, and $t\,\tau$ with
 100\% branching fraction,
 respectively, at 95 \% C.L. The previous results\cite{Khachatryan:2015vaa,Khachatryan:2015bsa} at 8 $\tev$ from the search of single leptoquark
 production are of order $0.65 \,\, \tev$ for final state $c\,\mu$. Taking a cue from these studies, we take $m_L > 1.5 \,\tev$ in our analysis.
 
 Similar to the leptoquarks, diquarks can be looked at the LHC through dijets in the final state. The recent studies at CMS on dijets' final states rules out  
 scalar diquarks of mass smaller than $6 \,\, \tev$. However, these limits are derived for $E_6$ diquark which couples with an up-type quark and a
 down-type quark\cite{Khachatryan:2015dcf}. These limits are very sensitive to the assumptions of decay branching fractions as well as the flavor dependent
 coupling strengths. Also, the diquark in the present work couples only to down-type quarks. This leads to a decrease in the flux factor 
 and hence the cross section
 and thereby the bounds on $m_S$ would be lower. Hence, we take $m_{S} \,\in \, [5:20] \, \tev$ in our analysis. 
 
 With this mass range of scalars, we randomly generate a sample of diquark couplings satisfying the constraints discussed in Sec. III.
 For $m_{S} \,\in
 \, [5:20]\, \tev$, the $B \to \pi K$ fit requires $Y_d^{12,13}$ to be greater than 0.1. Thus, we generate these couplings randomly in the range $[0.1:1]$. We fix 
 $Y_d^{23}$ of the order $10^{-2}$ and $Y_d^{33}$ is randomly generated in the range $ [10^{-4}:10^{-2}]$. The small value of 
 $Y_d^{33}$ is required to generate a small neutrino mass because the $Y_d^{33}$ coupling is always multiplied to the square of a 
 bottom quark mass when mass matrix, in the Eq. \ref{masnu}, is solved.
 For the remaining $Y_d^{ij}$, i.e, $Y_d^{11,12}$, we scan in the range $[10^{-5}:1]$. Except for $Y_d^{23}$, other diquark couplings are assumed complex. It 
 should be noted that the signs of the couplings are randomly assigned with equal probabilities being positive or negative in the whole calculation. 
 
\begin{figure*}[thb]
\centering
\begin{tabular}{ccc}
 \includegraphics[width=55mm]{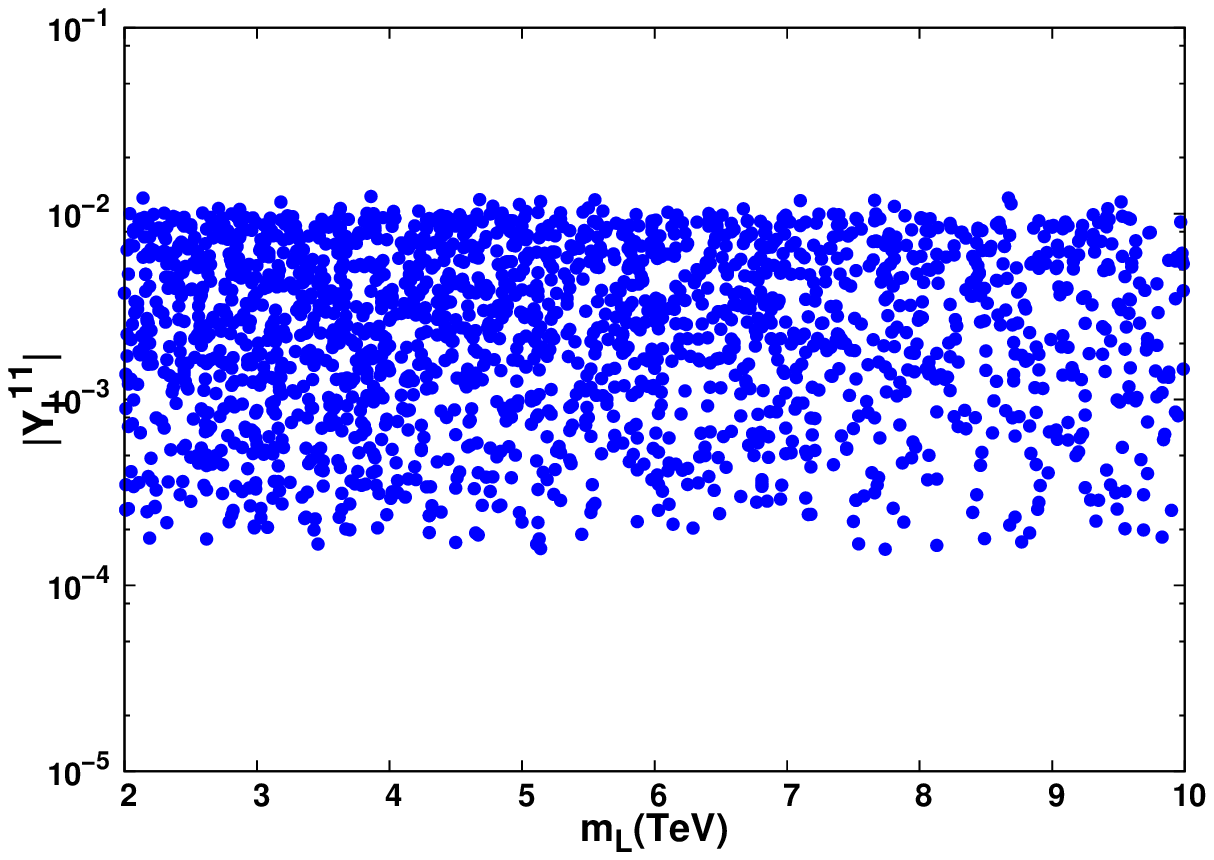} &
 \includegraphics[width=55mm]{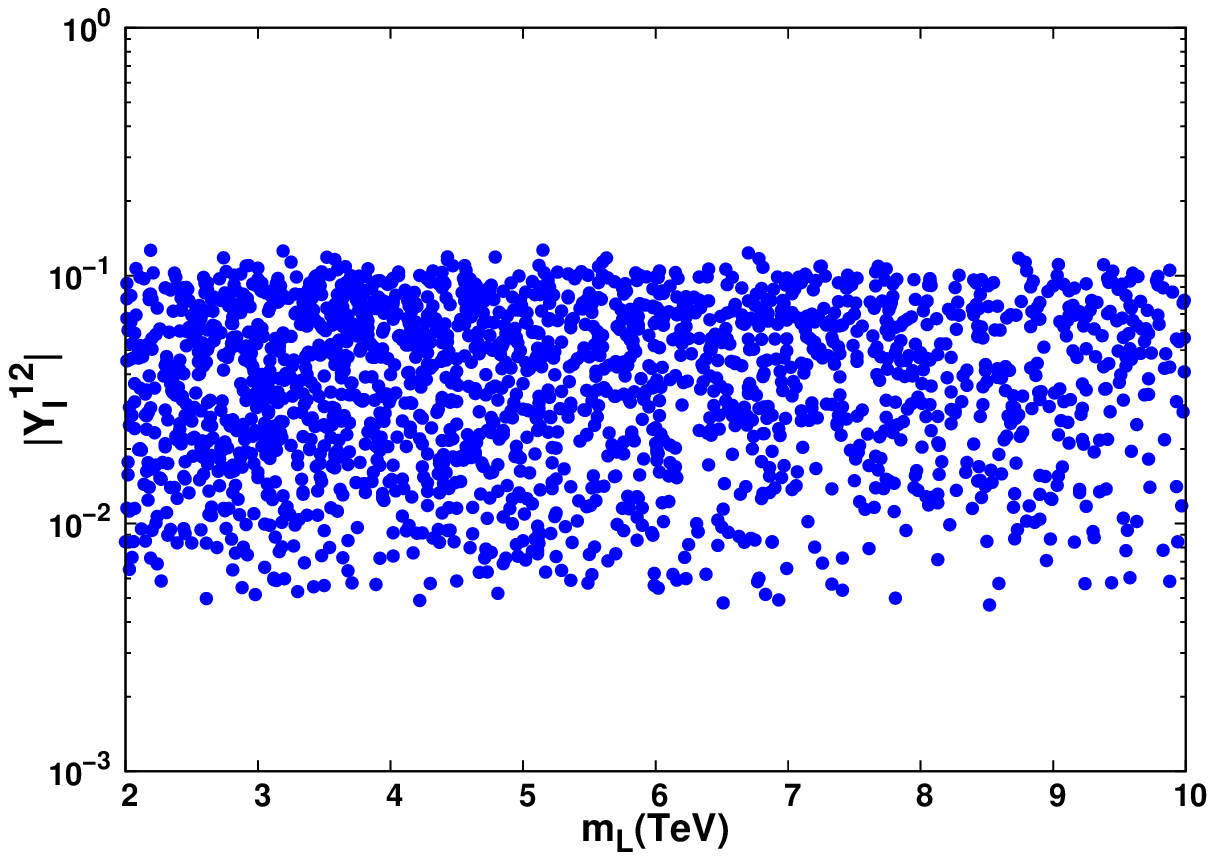} &
 \includegraphics[width=55mm]{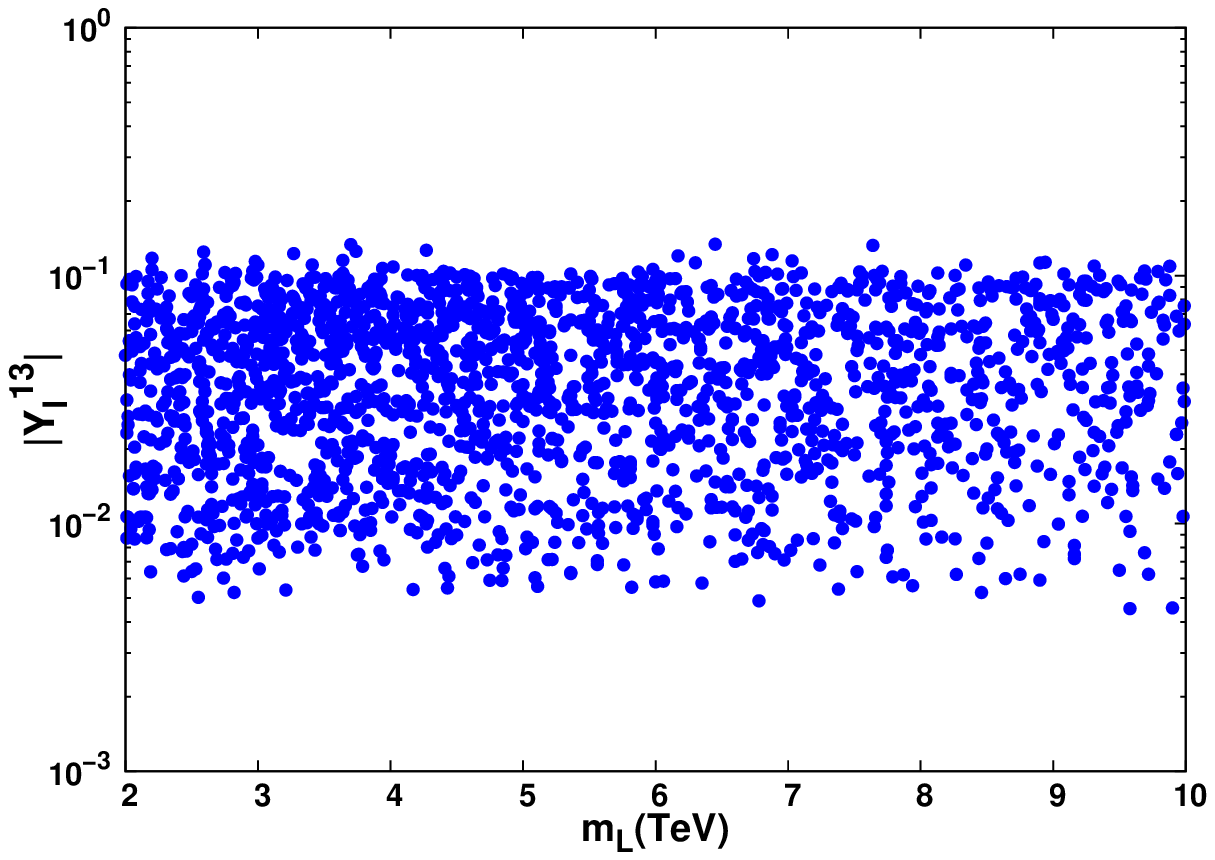} \\
 \includegraphics[width=55mm]{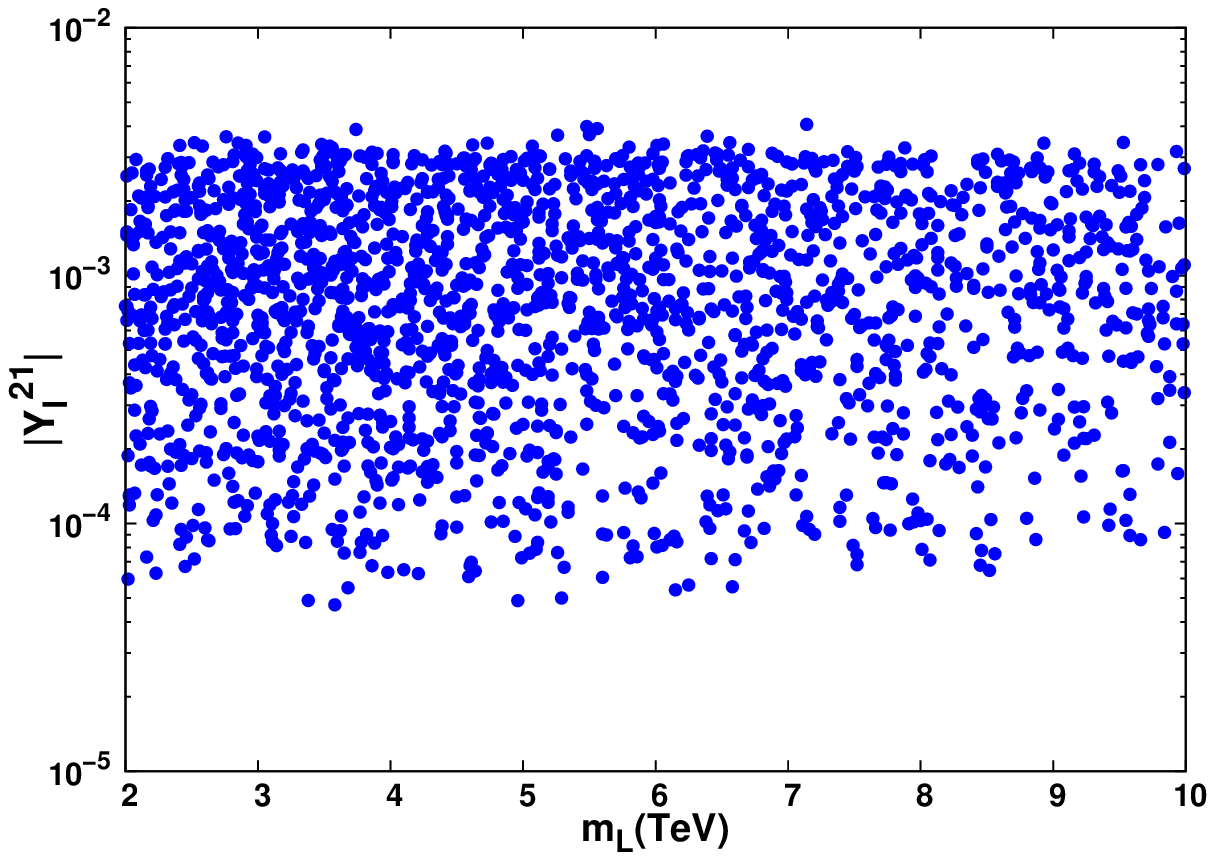} &
 \includegraphics[width=55mm]{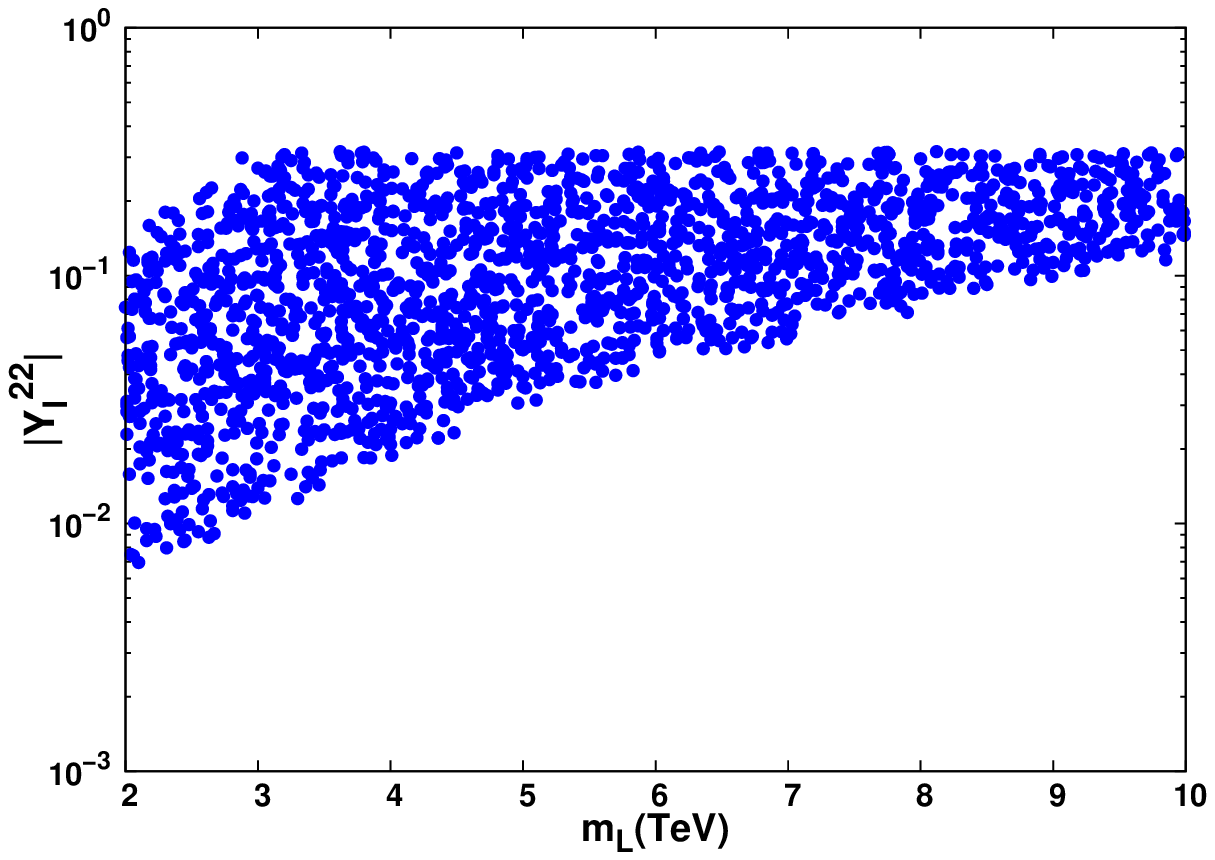} &
 \includegraphics[width=55mm]{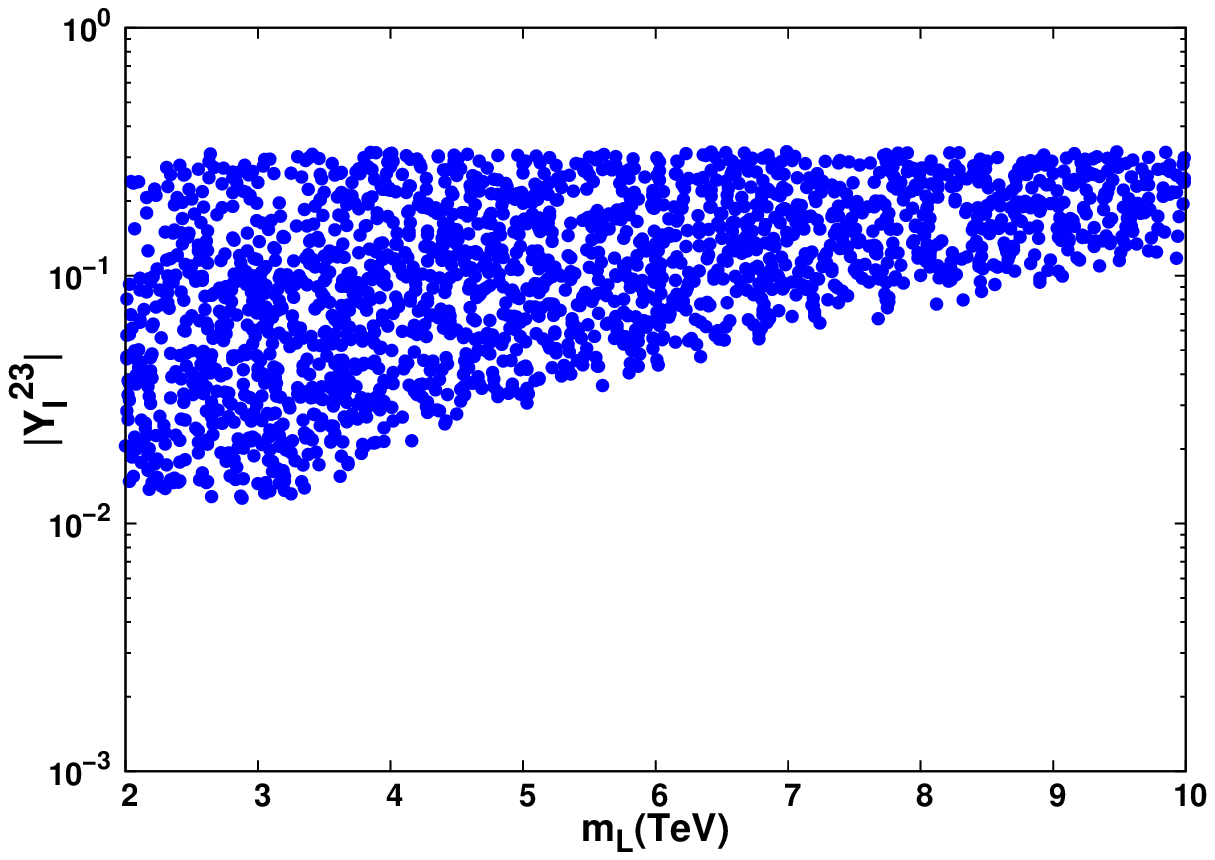} \\
 \includegraphics[width=55mm]{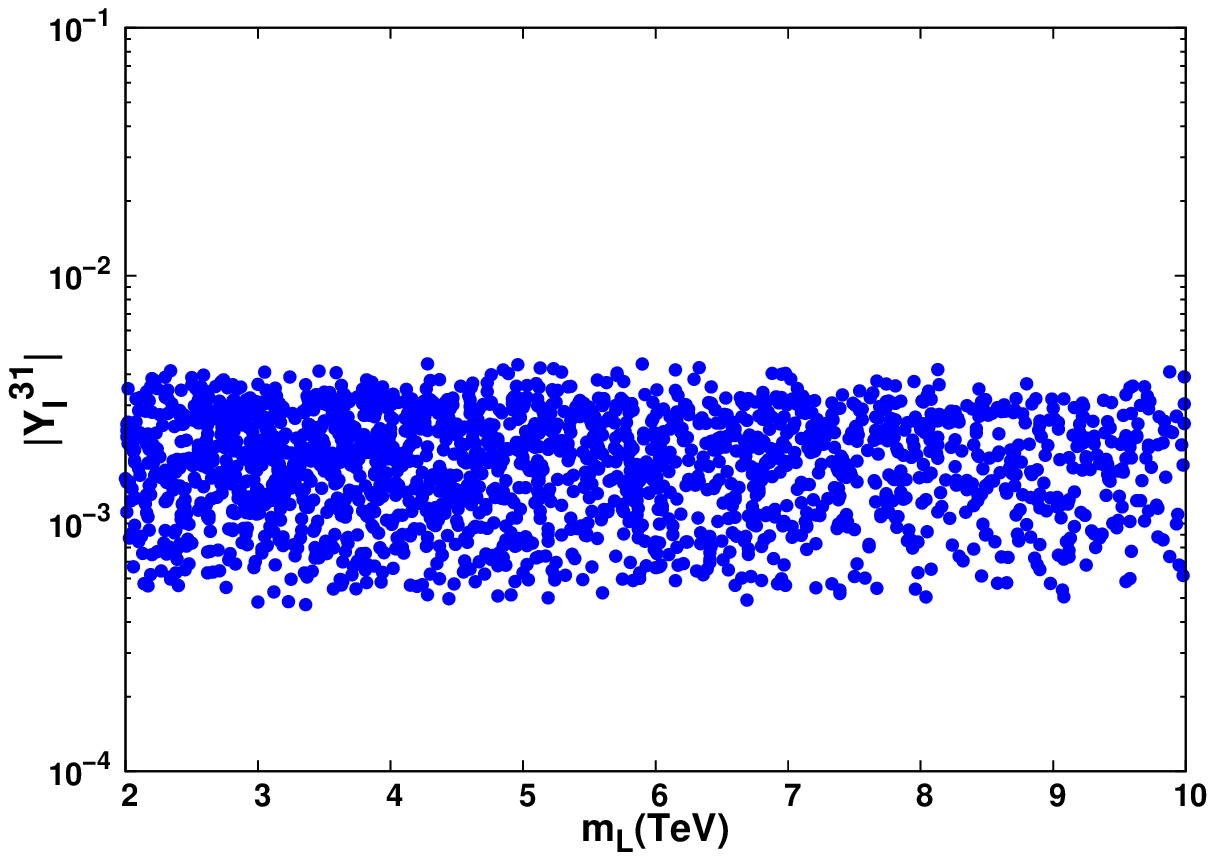} &
 \includegraphics[width=55mm]{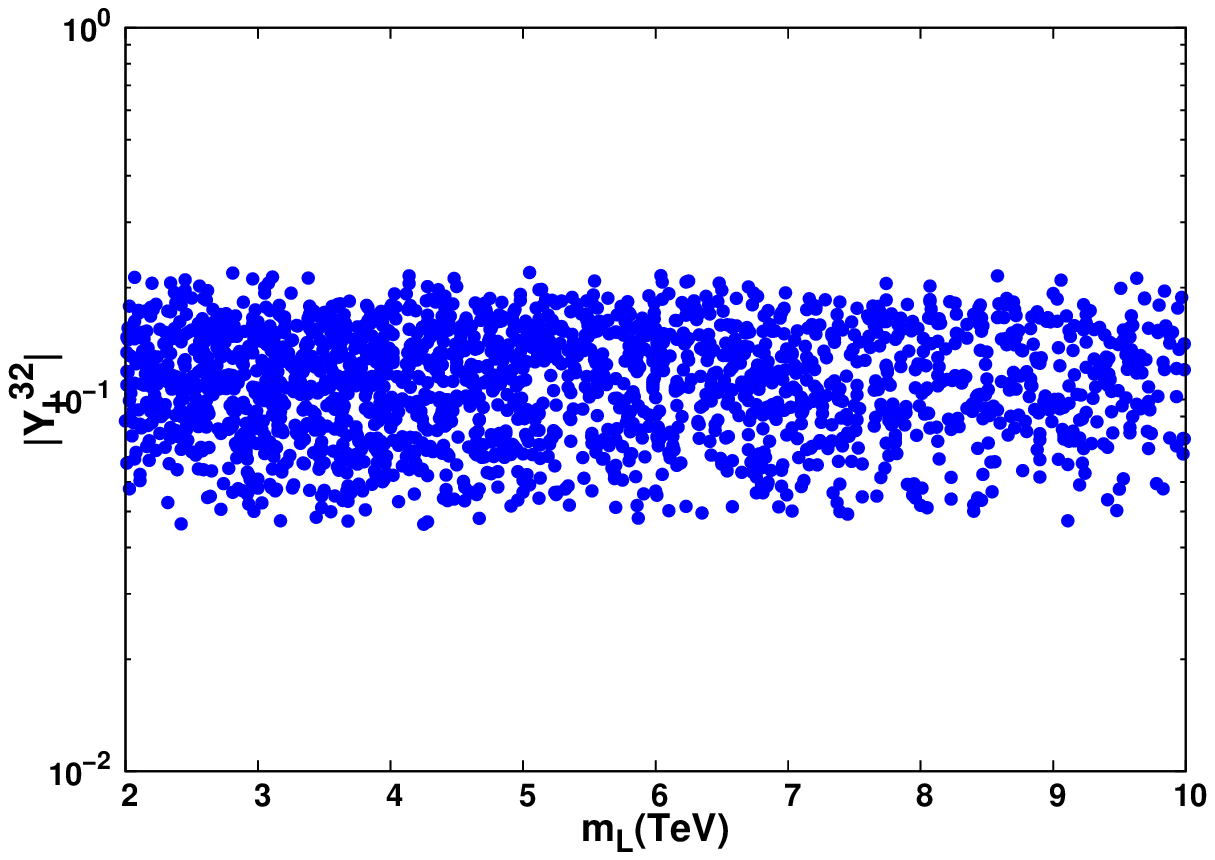} &
 \includegraphics[width=55mm]{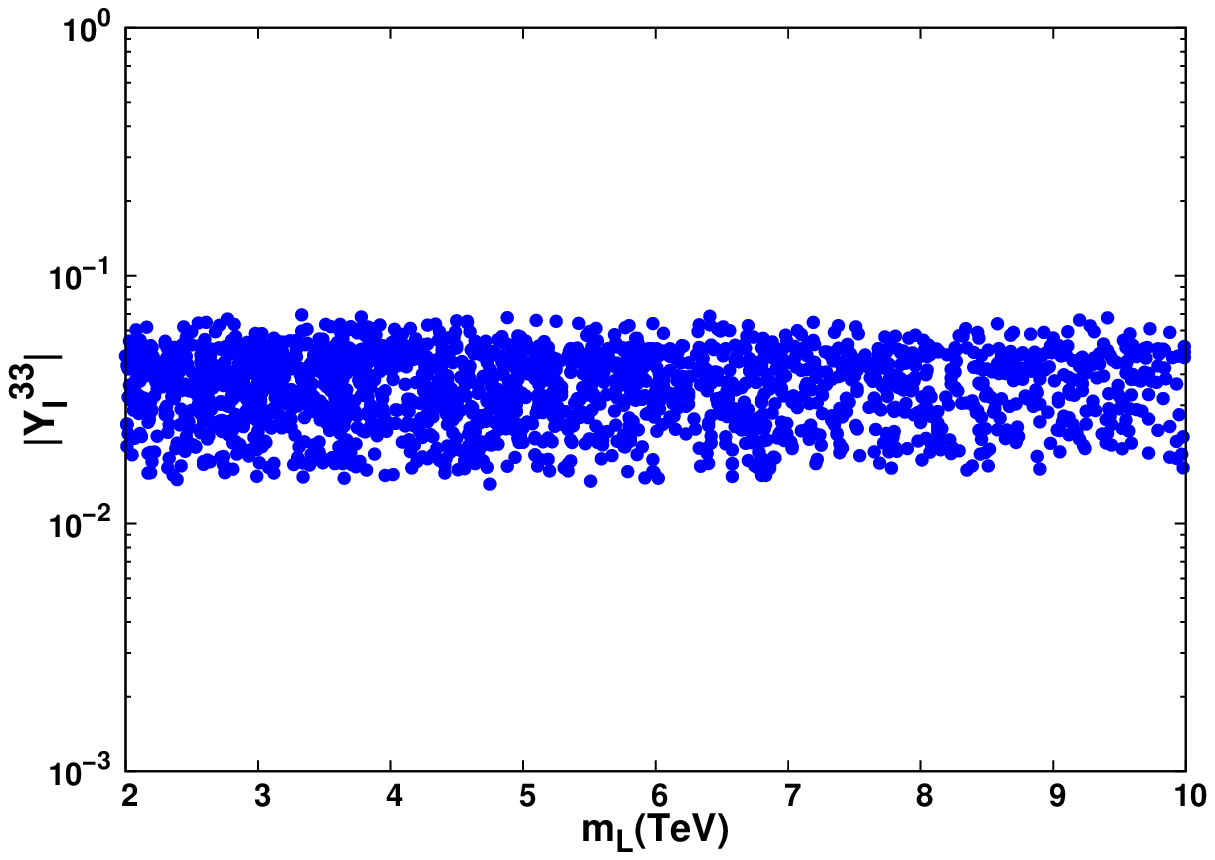} \\
\end{tabular}
\caption{\em Parameter space scan in $Y_l^{ij}$-$m_L$ plane.}
\label{Fig:LQ} 
\end{figure*}

 As for the leptoquark case, $Y_l^{2i}$ couplings(real) are generated randomly in the range $[10^{-5}:1]$. With the obtained sets of couplings, we calculate
 the strength of remaining leptoquark couplings, for randomly generated LQ mass, from Eq. \ref{masnu} to get the correct 
 neutrino masses. The symmetric neutrino mass matrix in the Eq. \ref{masnu} represents six independent equations as six 
 independent parameters (given in Table~\ref{tab:nu_exp}) that are obtained from the neutrino oscillation experiments.
 Throughout the analysis, we have kept Majorana phases to be 0, and have employed the 2$\sigma$ ranges for the neutrino 
 mixing parameters for normal hierarchy from Refs.\cite{Capozzi:2017ipn,PhysRevD.98.030001}. Finally, those sets of LQ
 couplings are selected that satisfy all of the constraints in Sec. \ref{sec:2}. The 
 results for the LQ couplings are given in Fig.\ref{Fig:LQ}.
 
 The pattern in the lower limit of the $Y_l^{22,23}$ coupling is mainly decided by $b\to s \ell^+\ell^-$ anomalies whereas the DQ couplings,
 $Y_d^{12/13}$, do not contribute 
 significantly to neutrino mass calculations and thereby leptoquark parameter space as $Y_d^{12/13}$ comes with the product of down and strange/bottom quark masses
 in Eq. \ref{masnu}, and the down quark mass is very small.
 
 We compare our results for leptoquark coupling with the results given in \cite{Guo:2017gxp} and \cite{Hiller:2017bzc} and find them consistent.
 A few benchmark points (BP) are given in Appendix A. For these BP, we present branching ratios for the rare decays in Table \ref{Tab:rare}
 following the calculations in Ref.~\cite{Giudice:2011ak}.  The branching ratios are rather small and it will be difficult to
 observe these decays in ongoing experiments.
 Our analysis shows that the $B$ anomalies and the neutrino masses can all be accommodated in a consistent framework. 
 
\begin{table*}[t]
\begin{center}
\begin{tabular}{|c|c|}
\hline\hline
$\delta m^2$  & $ 7.07-7.73 \times 10^{-5} \mathrm{eV}^2$  \\
$\sin^2 \theta_{12}$  &  $0.265-0.334$ \\
$|\Delta m^2|$& $2.454-2.606\times 10^{-3}\mathrm{eV}^2$  \\
$\sin^2 \theta_{13}$  	&  $0.0199-0.0231$\\
$\sin^2 \theta_{23}$ &	$0.395-0.470$ \\												
$\delta/\pi$      & 	$1.00-1.90$  \\
						\hline\hline
\end{tabular}
\end{center}
\caption{Neutrino data with $2\sigma$ deviation for normal hierarchy\cite{Capozzi:2017ipn,PhysRevD.98.030001}. }
\label{tab:nu_exp}
\end{table*}

\begin{table}[H]
\centering
 \begin{tabular}{|c|c|c|c|}
 \hline
  B.P & $\mathbf{\text{BR}(B^{\pm} \to \phi \pi^{\pm})}$ & $\mathbf{\text{BR}(B^{0} \to \phi \pi^{\circ})}$ & $\mathbf{\text{BR}(B^{0} \to \phi \phi)}$ \\
  \hline
  A & 1.45 $\times 10^{-10}$&7.2 $\times 10^{-11}$ & 1.45 $\times 10^{-12}$ \\
  B & 6.5 $\times 10^{-14}$&3.2 $\times 10^{-14}$ & 6.5 $\times 10^{-16}$ \\
  C & 1.19 $\times 10^{-12}$&5.95 $\times 10^{-13}$ & 1.19 $\times 10^{-14}$ \\
  \hline
 \end{tabular}
\caption{Branching ratios obtained with the couplings that can produce required neutrino mass and 
also satisfy the constraints coming from the $B \to \pi K$ puzzle.}
\label{Tab:rare}
\end{table}
\section{Conclusion} In conclusion we have discussed a unified framework to provide solutions to three problems. 
They are the anomalies in $\bsmumu$ measurements, nonleptonic
$\btopik$ decays, and the issue of generating neutrino masses and mixing. Our framework contained a scalar triplet leptoquark, 
a scalar color sextext diquark, and also, possibly, a color
antitriplet diquark. We considered several low energy as well as collider bounds on the leptoquark, diquark couplings and masses. For the leptoquarks these low energy observables included the $\bsll$ measurements. The solutions to the $\btopik$ puzzle
provided constraints on products of the diquark Yukawa couplings. We then checked that the correct neutrino masses and mixings were reproduced with the allowed
couplings of the leptoquarks and diquarks. We also predicted the branching ratios for a few rare $B$ decays whose observations could signal the existence of diquarks. However, we found the branching ratios of these decays to be unobservably small.

\section{Acknowledgments}: We thank Ernest Ma for suggesting this problem. 
D.S. acknowledges the computing facility provided under SERB, India's project Grant No. 
EMR/2016/002286 and thanks UGC-CSIR, India, for financial assistance.

\appendix
\section{Some Useful Expressions}
In this Appendix, we give some useful expressions and calculations that could be useful while reading the paper 
\bea
\psi^c &= &C \overline{\psi}^T \nonumber,\\
\overline{\psi}^c & = & (\psi^c)^\dagger \gamma^0 = -\psi^T C^{-1}, \nonumber\\
(\gamma^\mu)^T &= & -C^{-1} \gamma^\mu C, \nonumber\\
C^{-1} & = & C^\dagger, \
\eea
\bea
H & = & Y_d^{ij} \overline{d}^c_{i \alpha} P_R d_{j \beta} S^{\alpha \beta},  \nonumber\\
H^\dagger & = &   Y_d^{ij*} {d}_{j \beta}^\dagger P_R(-d_{i \alpha} ^TC^{-1})^\dagger S^{*\alpha \beta}  \nonumber\\
& = & - Y_d^{ij*} \overline{d}_{j \beta} P_L( \gamma^0 C d_{i \alpha} ^*)S^{*\alpha \beta}.  \nonumber\\
\eea
Integrating out diquark
\bea
H_{eff} & = &  - Y_d^{13} \overline{d}^c_{ \alpha} P_R b_{\beta} S^{\alpha \beta} \otimes  Y_d^{12*} {s}_{ \beta}^\dagger P_R(-d_{\alpha} ^TC^{-1})^\dagger S^{*\alpha \beta} \nonumber\\
& = & -\frac{Y_d^{13}  Y_d^{12*}}{m_S^2}
 \overline{d}^c_{ \alpha} P_R b_{\beta}    \overline{s}_{ \beta} P_L( \gamma^0 C d_{\alpha} ^*) \nonumber\\
 & = &   
   \frac{Y_d^{13}  Y_d^{12*}}{2m_S^2}   \overline{d}^c_{ \alpha} \gamma^\mu P_L  ( \gamma^0 C d_{\alpha} ^*)      \overline{ {s}}_{ \beta} \gamma_\mu P_R  b_{\beta}      \nonumber\\
  & = &   \frac{Y_d^{13}  Y_d^{12*}}{2m_S^2}   \overline{s}_{ \beta} \gamma_\mu P_R b_{\beta}    \left[- {d}^T_{ \alpha} C^{-1} \gamma^\mu P_L  ( \gamma^0 C d_{\alpha} ^*) \right] \nonumber\\   
   & = & -\frac{Y_d^{13}  Y_d^{12*}}{2m_S^2} \overline{s}_{ \beta} \gamma_\mu P_R  b_{\beta}    \left[ {d}^T_{ \alpha} \gamma^{\mu T} P_L^T  ( \gamma^{0T}  d_{\alpha} ^*)\right]  \nonumber\\
    & = &  - \frac{Y_d^{13}  Y_d^{12*}}{2m_S^2}   \overline{s}_{ \beta} \gamma_\mu P_R  b_{\beta}    \left[ d_{\alpha} ^\dagger \gamma^0 P_L \gamma^\mu    {d}_{ \alpha}  \right]^T  \nonumber
    \\
  & = & \frac{Y_d^{13}  Y_d^{12*}}{2m_S^2}  \overline{s}_{ \beta} \gamma_\mu P_R  b_{\beta}     \overline{d_{\alpha} } \gamma^\mu P_R     {d}_{ \alpha}    \ 
   \eea
Because $S^{\alpha \beta}$ is symmetric/antisymmetric there is an additional factor of 2. In other words $S^{12}$ can contract with $S^{12}$ and $S^{21}$.
\section{Benchmark Points}
Here we give the benchmark points satisfying the $B$ anomalies observations and explaining the neutrino mass.
\begin{itemize}
\item BP A: 
\[ m_L = 3.5 \tev, \qquad m_S = 5 \tev \]
\beq
Y_l = \left(
\begin{array}{ccc}
\dis 1.40\times 10^{-4}+i3.24\times 10^{-4}&	5.02\times 10^{-3}+i8.9\times 10^{-3}&	3.7\times 10^{-3}+i3.26\times 10^{-2}\\[1.5ex]
1.37\times 10^{-3}+i2.83\times 10^{-4}&	1.81\times 10^{-1}&2.44\times 10^{-2}\\[1.5ex]
5.03\times 10^{-4}+i3.12\times 10^{-3}&1.4\times 10^{-1}+i 3.31\times 10^{-2}&1.1\times 10^{-2}+i4.5\times 10^{-2}
         \end{array}
         \right)\ .         
\eeq
\beq
Y_d = \left(
\begin{array}{ccc}
\dis 1.68\times 10^{-4} &	4.6\times 10^{-1}+i1.22\times 10^{-1} &	4.64 \times 10^{-1}+i 1.3\times 10^{-2} \\[1.5ex]
4.6\times 10^{-1}+i1.22\times 10^{-1}&	2\times 10^{-1}&	0.01 \\[1.5ex]
 4.64 \times 10^{-1}+i 1.3\times 10^{-2} &		0.01&	-1.42\times 10^{-4} +i 2.5 \times 10^{-4}
         \end{array}
         \right)\ .
\eeq
\[(M_\nu)_{ee}= 4.53 \times 10^{-3} \ev\]

\item BP B:
\[ m_L = 7.5 \tev, \qquad m_S = 6 \tev \]
\beq
Y_l = \left(
\begin{array}{ccc}
\dis 1.03\times 10^{-4}+i7.8\times 10^{-3}&	8.2\times 10^{-3}+i1.2\times 10^{-2}&	1.87\times 10^{-2}+i1.11 \times 10^{-2}\\[1.5ex]
1.32\times 10^{-3}+i3.2\times 10^{-4}&	2.15\times 10^{-1}	&9.5\times 10^{-2}\\[1.5ex]
7.56\times 10^{-4}+i1.91\times 10^{-3}&	1.23\times 10^{-1}+i1.25\times 10^{-1}&	3.2\times 10^{-2}+i1.51\times 10^{-2}
         \end{array}
         \right)\ .         
\eeq
\beq
Y_d = \left(
\begin{array}{ccc}
\dis 1.38\times 10^{-4}  &	6.28\times 10^{-2} +i3.6\times 10^{-1}&	5.1\times 10^{-1} +i2.12\times 10^{-2} \\[1.5ex]
6.28\times 10^{-2} +i3.6\times 10^{-1} &	1.8\times 10^{-1} & 0.01\\[1.5ex]
5.1\times 10^{-1} +i2.12\times 10^{-2} & 0.01 & -1.4\times10^{-3}+i3\times 10^{-4}
\end{array}
         \right)\ .
\eeq
\[(M_\nu)_{ee}= 1.55 \times 10^{-3} \ev\]

\item BP C: 
\[ m_L = 5.0 \tev, \qquad m_S = 7.5 \tev \]
\beq
Y_l = \left(
\begin{array}{ccc}
\dis 5.1\times 10^{-3}+i2.63\times 10^{-4}&	4.6\times 10^{-2}+i5.2\times 10^{-2}&	3.3\times 10^{-3}+i1.1 \times 10^{-2}\\[1.5ex]
7.26\times 10^{-4}+i1.55\times 10^{-3}&	2.42\times 10^{-1}	&4.3\times 10^{-2}\\[1.5ex]
1.57\times 10^{-3}+i1.64\times 10^{-3}&	1.24\times 10^{-1}+i1.06\times 10^{-1}&	1.32\times 10^{-2}+i1.0\times 10^{-2}
         \end{array}
         \right)\ .         
\eeq
\beq
Y_d = \left(
\begin{array}{ccc}
\dis 1.2\times 10^{-4} &	3.04\times 10^{-1} +i7.3\times 10^{-1}&	5.1\times 10^{-1} +i1.79\times 10^{-1} \\[1.5ex]
3.04\times 10^{-1} +i7.3\times 10^{-1} &	7.2\times 10^{-1} &	0.01\\[1.5ex]
5.1\times 10^{-1} +i1.79\times 10^{-1} & 0.01 & -1.43\times 10^{-2}-i5.11\times 10^{-3}
\end{array}
         \right)\ .
\eeq
\[(M_\nu)_{ee}= 1.01 \times 10^{-3} \ev\]
\end{itemize}

\bibliographystyle{apsrev4-1}
\bibliography{reference}

\end{document}